\documentclass[prl,twocolumn]{revtex4}
\usepackage{amsfonts}
\usepackage{amsmath}
\usepackage{amssymb}
\usepackage{graphicx}

\begin{document}

\title{Hot Spots and Pseudogaps for Hole- and Electron-Doped
High-Temperature Superconductors. }
\author{David S\'{e}n\'{e}chal$^{1}$ and A.-M.S. Tremblay$^{1,2}$}
\affiliation{$^{1}$D\'{e}partement de physique and Regroupement qu\'{e}b\'{e}cois sur les
mat\'{e}riaux de pointe, $^{2}$Institut canadien de recherches avanc\'{e}es,
Universit\'{e} de Sherbrooke, Sherbrooke, Qu\'{e}bec, Canada, J1K 2R1}
\date{August 2003}

\begin{abstract}
Using cluster perturbation theory, it is shown that the spectral weight and
pseudogap observed at the Fermi energy in recent Angle Resolved
Photoemission Spectroscopy (ARPES) of both electron and hole-doped
high-temperature superconductors find their natural explanation within the $%
t$-$t^{\prime }$-$t^{\prime \prime }$-$U$ Hubbard model in two dimensions. The
value of the interaction $U$ needed to explain the experiments for
electron-doped systems at optimal doping is in the weak to intermediate
coupling regime where the $t-J$ model is inappropriate. At strong coupling,
short-range correlations suffice to create a pseudogap but at weak coupling
long correlation lengths associated with the antiferromagnetic wave vector
are necessary.
\end{abstract}

\maketitle


Deep insight into the nature of strongly correlated electron materials, such
as high temperature superconductors, has emerged in the last few years from
both experiment and theory. On the experimental side, ARPES~\cite%
{Damascelli03} and scanning-tunneling experiments~\cite{Davis02} provide us
with detailed information on the nature of single-particle states. This
information must be explained by theory if we are to understand correlated
materials. For example, contrary to one of the central tenets of Fermi
liquid theory, sharp zero-energy excitations are not enclosing a definite
volume in the Brillouin zone. Certain directions are almost completely
gapped while others are not. This is the famous pseudogap problem that has
been the focus of much attention in the field~\cite{Timusk99}.

On the theoretical side, Dynamical Mean-Field Theory (DMFT)~\cite{Georges96}
has allowed us to understand the evolution of single-particle states during
the interaction-induced (Mott) transition between metallic and insulating
states~\cite{Imada98} (parent compounds of high-temperature superconductors
are Mott insulators). Generalizations of DMFT, such as the Dynamical Cluster
Approximation~(DCA) \cite{Maier00} and Cellular-DMFT~\cite{Bolech03} are
however necessary to take into account the momentum dependence of the
self-energy that is neglected in DMFT and is clearly apparent in ARPES
experiments~\cite{Damascelli03}. Up to now, these calculations have been
restricted to hole-doped systems and small system sizes or to the perfectly
nested case. The nature of single-particle excitations, and in particular
the pseudogap in cuprate superconductors, is thus still an open theoretical
problem.

Without any assumption about the nature of the ground state, we show that
the Hubbard model with fixed first-, second- and third-neighbor hopping ($t$
, $t^{\prime }$ and $t^{\prime \prime }$) accounts for the strikingly
different locations of low energy excitations observed experimentally in
hole- and electron-doped cuprate superconductors~\cite{Armitage01, Ronning03}.
At zero doping we have a Mott insulator with a large $U$. By contrast with
previous attempts to obtain a unified model~\cite{Kim98}, we will see that
the interaction strength $U$ varies as one moves from the hole-doped to the
electron-doped systems. That parameter should be at least of the order of
the bandwidth for hole-doped systems. In this case the pseudogap is
controlled mainly by Mott Physics with short-range correlations. The
situation is similar for underdoping with electrons. As we approach the
optimally-doped electron case, the pseudogap occurs at a smaller
coupling where Mott Physics is not essential. Long correlation lengths then
play an essential role in creating the pseudogap whereas in the strong
coupling case they are not necessary for the pseudogap to appear. These
results give insight into two different mechanisms for the
pseudogap phenomenon and into the nature of the breakdown of Fermi-liquid
theory in these systems. We also gain insight into the appropriate
microscopic model of high-temperature superconductors.

\begin{figure}[tbp]
\centerline {\includegraphics[width=8cm]{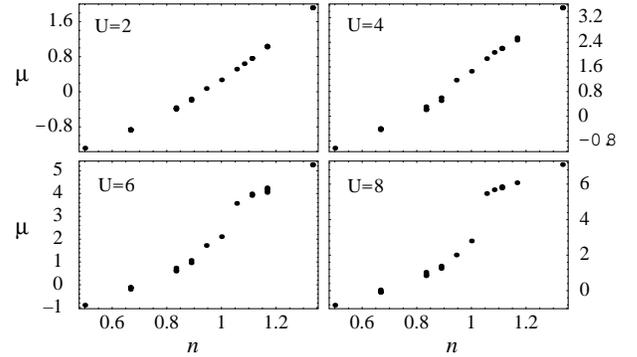}}
\caption{Chemical potential calculated at various dopings using CPT, in
units of $t$, the NN hopping. For this figure only, $t^{\prime }=-0.4t$ and $%
t^{\prime \prime }=0$.}
\label{FIG:mu_vs_n}
\end{figure}

\begin{figure}[ptb]
\centerline {\includegraphics[width=8.5cm]{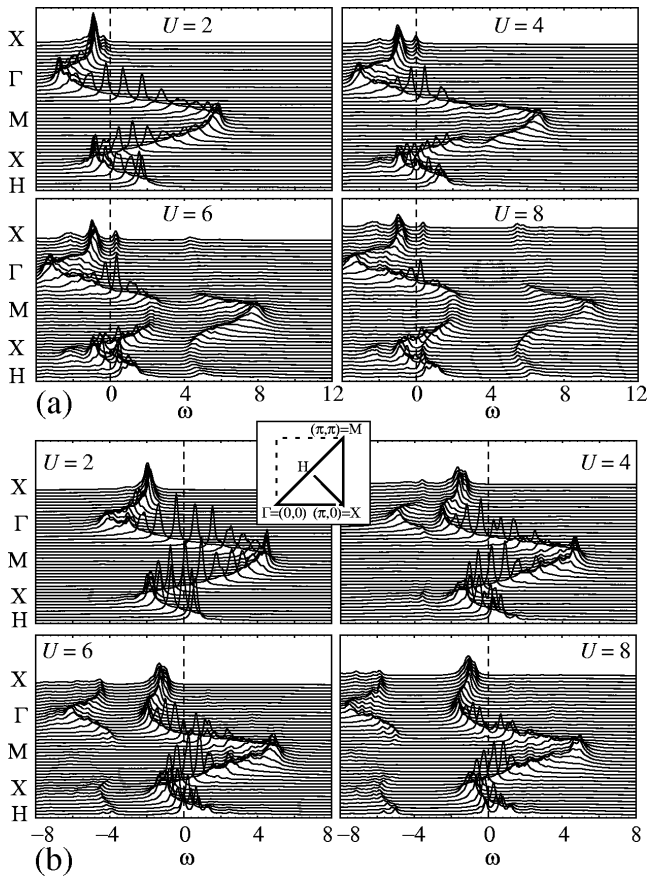}} \vglue -2mm
\caption{Single particle spectral weight, as a function of energy $\protect%
\omega$ in units of $t$, for wavevectors along the high-symmetry directions
shown in the inset. (a): CPT calculations on a $3\times4$ cluster with 10
electrons (17\% hole doped). (b): the same, with 14 electrons (17\% electron
doped). In all cases we use $t^{\prime}=-0.3t$ and $t^{\prime\prime}=0.2t$.
A Lorentzian broadening $\protect\eta=0.12t$ is used to reveal the otherwise
delta peaks. }
\label{FIG:spectre}
\end{figure}

\paragraph{Model and methodology.}

We study the square lattice Hubbard model with on-site Coulomb
repulsion $U$. We set the
first-neighbor hopping $t$ to unity, and introduce second-neighbor
(diagonal) hopping $t^{\prime }=-0.3t$ and third-neighbor hopping $t^{\prime
\prime }=0.2t$, as suggested by band structure calculations~\cite{Andersen95}%
. The diagonal hopping $t^{\prime }$ is a key ingredient to understand the
Physics, even though its precise value can vary slightly between different
compounds. It frustrates antiferromagnetic (AFM) order and removes
particle-hole symmetry, thereby also allowing the AFM zone boundary to cross
the Fermi surface. The third-neighbor hopping $t^{\prime \prime }$ makes the
Fermi surface slightly bulge away from the intersection with the AFM zone
boundary, as observed experimentally~\cite{Armitage01}, and makes low-energy
excitations more stable along the diagonal of the Brillouin zone.

We use Cluster Perturbation Theory~\cite{Senechal00} (CPT) to gain insight
into the single-particle states of the Hubbard model and their relation to
cuprate superconductors. The method can reproduce the spin-charge separation
of one dimensional systems~\cite{Senechal00} as well as the dispersion
relations obtained in the large $U$ limit. It reduces to the exact result at
$U=0$ and in the atomic limit ($t_{ij}=0$). It is based on exact
diagonalizations of finite clusters that are coupled through strong-coupling
perturbation theory. It basically amounts to replacing the exact self-energy
by that of the cluster only~\cite{Gros94}. The Green function calculated by
CPT is made up of a set of discrete poles, like in ordinary exact
diagonalizations, except that (i) more poles have substantial weight and
(ii) they disperse continuously with wavevector, allowing for clear momentum
distribution curves. The results presented here were calculated on $12$-site
rectangular clusters. The resulting Green function is averaged over the ($%
3\times 4$) and ($4\times 3$) clusters to recover the original symmetry of
the lattice. We checked that the main features are the same when using
clusters of different shapes. Our finite energy resolution, of about $0.12t$%
, does not allow us to resolve effects related to superconductivity. We
compare with ARPES experiments of similar resolution.


\paragraph{The Mott transition.}

We begin in Fig.~\ref{FIG:mu_vs_n} with a plot of the chemical potential $%
\mu $ as a function of doping for various values of the interaction
strength. The different results in this figure are obtained from clusters of
different sizes ranging from $4$ to $13$ sites with varying geometry. The
smooth behavior of the function away from half-filling shows that the
cluster sizes are large enough to provide reliable results. There is a jump
in $\mu $ when $U$ is large enough, namely above $U=6t$ roughly. The jump in
$\mu $ does not follow from a long-range ordered ground state since the
basic clusters are finite. It is instead a clear manifestation of the Mott
phenomenon.

Fig.~\ref{FIG:spectre} displays the single-particle spectral weight $A(
\mathbf{k},\omega )$ as a function of energy for wave vectors $\mathbf{k}$
along the high-symmetry directions shown in the inset. Only the $\omega <0$
domain of $A(\mathbf{k},\omega )$ is accessible to ARPES. Fig.~\ref%
{FIG:spectre}(a) illustrates the effect of increasing interaction strength
on a near optimally hole-doped system while Fig.~\ref{FIG:spectre}(b) does
the same in the electron-doped case. Clearly, there is a range of
frequencies where $A(\mathbf{k},\omega )=0$ for \textit{all} wave vectors.
This is the Mott gap. At finite doping it always opens up away from zero
energy when $U$ is sufficiently large. In the electron-doped case, $\omega
=0 $ is in the upper Hubbard band. The lower Hubbard band is at negative
energies, as has been observed in ARPES~\cite{Armitage02}. The overall
narrowing of the band just below the Fermi level is more important in the
electron-doped case. Also, the shape of the dispersion is different from
that obtained in a mean-field AFM state~\cite{Kusko02} and there is no clear
doubling of the dispersion relation of the type that occurs in one dimension
when there is spin-charge separation.

\begin{figure}[ptb]
\centerline {\includegraphics[width=8cm]{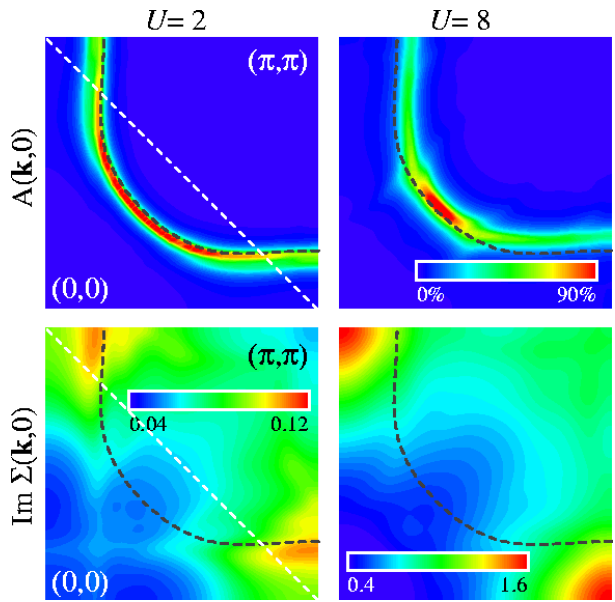}}
\vglue-2mm
\caption{(color) Top: Intensity plot of the spectral function at the Fermi
level, in the first quadrant of the Brillouin zone, for a 17\% hole-doped
system (10 electrons on a $3\times4$ cluster). Here $t^{\prime}=-0.3t$ and $
t^{\prime\prime}=0.2t$ (the gray dashed line is the non-interacting Fermi
surface). Bottom: Imaginary part of the self-energy (in units of $t$)
corresponding to the same parameters as the top plot. A Lorentzian
broadening is used: $\protect\eta=0.12t$ (top) and $\protect\eta=0.4t$
(bottom).}
\label{FIG:mdcH}
\end{figure}


\paragraph{Fermi surface plots and pseudogap.}

We now move to the main point of our paper, namely the pseudogap and hot
spots. The top two panels of Fig.~\ref{FIG:mdcH} for a 17\% hole-doped
system represent the strength of $A(\mathbf{k},\omega )$ at $\omega =0$. As
a function of interaction strength, the intensity disappears gradually near
the $(\pi ,0)$ and $(0,\pi )$ points, leaving zero-energy excitations only
near the diagonal. Large values of $U$ $\left( U>8t\right) $ seem necessary
to reproduce the experimentally observed spectral function of hole-doped
systems~\cite{Ronning03}, even more so on a 11\% doped system (not shown).
The lower panels show the imaginary part of the self-energy (or scattering
rate) corresponding to the momentum-dispersion curve right above. For $U=2t$%
, the self-energy is very small overall, but has a maximum along the Fermi
surface where the Fermi velocity is smallest (density of states largest);
this illustrates how we depart from the Fermi liquid picture (in which $%
\Sigma ^{\prime \prime }(0)=0$) as we move towards intermediate coupling. At
$U=8t$, the scattering rate is much larger and affects larger regions
separated by roughly $(\pi ,\pi )$. In all cases a higher scattering rate
leads to removal of spectral weight.

\begin{figure}[ptb]
\centerline{\includegraphics[width=8cm]{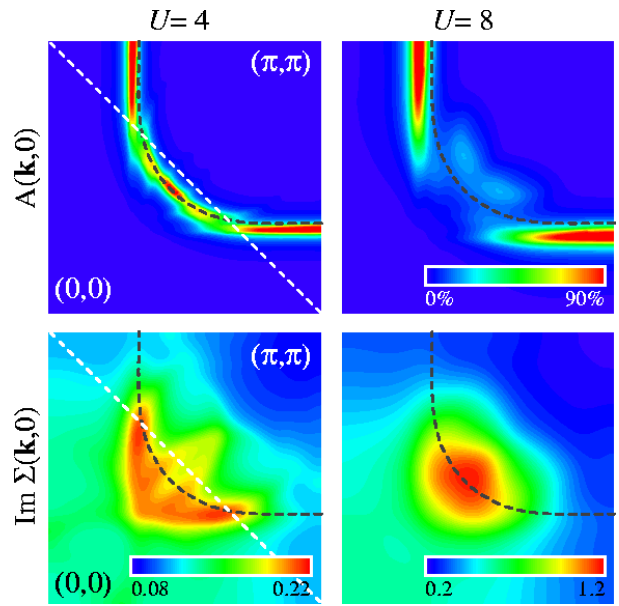}}
\vglue-2mm
\caption{(color) Same as Fig.~\protect\ref{FIG:mdcH}, but for an
electron-doped system (14 electrons on a $3\times4$ cluster). The white
dashed line on the left panels is the AFM zone boundary, showing the
coincidence of hot spots with the intersection of this line with the Fermi
surface.}
\label{FIG:mdcE}
\end{figure}

The electron-doped case is displayed in Fig.~\ref{FIG:mdcE} for $17\%$
doping. At moderate $U$, the spectral intensity drops only at the
intersection of the AFM Brillouin zone with the Fermi surface. However, for
larger $U$, only the neighborhoods of $(\pi ,0)$ and $(0,\pi )$ survive. The
latter situation is analogous to that observed by ARPES in electron
underdoped systems and can be reproduced by calculations (not shown) with $U$
large at $11\%$ doping. At optimal doping however, ARPES results~\cite%
{Armitage02} look instead qualitatively like the upper left panel of Fig.~%
\ref{FIG:mdcE}.


\paragraph{Hot spots and pseudogap.}

\begin{figure}[ptb]
\centerline {\includegraphics[width=8cm]{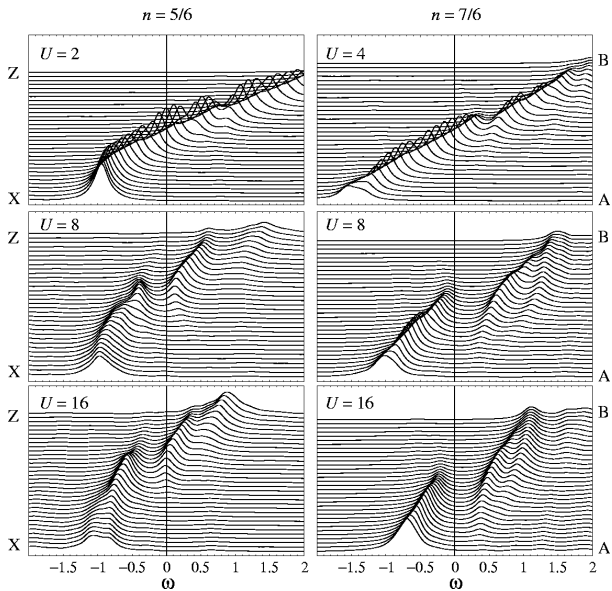}} \vglue -2mm
\caption{Left: Spectral function for the hole doped system illustrated in
Figs.~\protect\ref{FIG:spectre}a and \protect\ref{FIG:mdcH} plotted as a
function of energy, for wavevectors along the direction $X=(\protect\pi,0)$
to $Z=(\protect\pi,\protect\pi/2)$. At $U=2t$ (top), a depression in the
spectral function is visible slightly away from $\protect\omega=0$, while
the pseudogap is fully opened at $U=8t$ (middle). Right: Spectral function
for the electron-doped system illustrated in Figs \protect\ref{FIG:spectre}b
and \protect\ref{FIG:mdcE}, plotted along the diagonal of the Brillouin
zone, from $A=(0.3\protect\pi,0.3\protect\pi)$ to $B=(0.7\protect\pi,0.7
\protect\pi)$. The results for the experimentally relevant electron
underdoped system are similar.}
\label{FIG:pseudo}
\end{figure}

The Fermi-surface points where the intensity decreases (Fig.~\ref{FIG:mdcE})
are called hot spots. However, a pseudogap is characterized not
only by lower intensity at the Fermi energy, but also by a dispersive peak
that stops short of crossing the Fermi surface. This experimentally
well-known phenomenon is illustrated on Fig.~\ref{FIG:pseudo}, which shows
energy dispersion curves for wave vectors along the $(\pi ,0)-(\pi ,\pi /2)$
stretch in the hole-doped case (left) and along the diagonal in the
electron-doped case (right). For small values of $U$, a well-defined
quasiparticle exists at the Fermi level ($\omega =0$). At stronger coupling (%
$U=8t$), a pseudogap comparable to experimental observation is clearly
visible at the Fermi level.

\paragraph{Discussion: strong- and weak-coupling pseudogaps.}

As in previous studies, the strong-coupling pseudogap \cite{Phillips03} is
concomitant with the Mott gap but is clearly distinct from the latter. The
Mott gap is a purely local (on-site) phenomenon that occurs for all wave
vectors and is not tied to $\omega=0$. By contrast, the pseudogap occurs around $\omega=0$ and only in regions of the Fermi surface
that are connected to other such regions by
wave vectors that have a broad spread of radius $\delta$ around $(\pi,\pi)$.
The difference in the location of the pseudogap
between hole- and electron underdoped cuprates follows by simply finding which
points of the Fermi surface can be connected by $(\pi,\pi)$, within $\delta $,
to other Fermi surface points.

Despite the importance of $(\pi,\pi)$, the strong-coupling  pseudogap is
not caused by long-range AFM correlations.  Indeed, (a) Our lattices do
not exhibit long-range order (b) We verified  that the results are not
very sensitive to $t^{\prime }$ (frustration) (c)  Fig.~\ref{FIG:pseudo}
shows that at $U>8t$ the pseudogap is of order $t$,  only weakly
dependent on $U$ and does not scale as the antiferromagnetic  coupling
$J=4t^{2}/U$, in contrast with previous studies~\cite%
{Phillips03,Haule02}. This pseudogap would therefore persist in the $U\to\infty$
limit of the Hubbard model, where hopping
between sites is constrained by the impossibility of double occupancy
and where $t$ is the relevant energy scale. For a case
where it is possible to study the size  dependence of the
strong-coupling pseudogap at fixed doping, we verified  that the results
are size independent, suggesting again the short-range  nature of the
phenomenon.  Longer range
correlations at the AFM wave vector might only reinforce the strong-coupling pseudogap that
already exists in the  presence of short-range correlations. The location of this
strong-coupling pseudogap, in  both electron- and hole-doped cases,
coincides with the predictions of the  umklapp mechanism
\cite{FurukawaRice}, which does not need long-range correlations.
However, a proper strong-coupling extension of the umklapp mechanism is
still needed.

Signs of a pseudogap also occur at weak coupling. This is illustrated by
the hot spots that are visible in the electron-doped case at $U=4t$ in Fig.~\ref{FIG:mdcE}, upper panel. Contrary to the $U=8t$ case, these hot spots (a) are located precisely at the intersection with the AFM zone boundary (b) they generally correspond to a cluster shape dependent depression in $A(\mathbf{k},\omega )$ and not to a genuine pseudogap. We attribute these results to the short correlation lengths (limited to the cluster size) in CPT and conclude that we are seeing the onset of the true
pseudogap that, as expected from the presence of true gaps in the itinerant antiferromagnet~\cite{Markiewicz}, would be induced by large AFM correlation lengths~\cite{Kyung02}. We find, as in Ref.~\cite{Kyung02}, that the interaction strength $U$ cannot be larger than $U\approx 6t$ to preserve this kind of pseudogap where $\omega=0$ excitations persist near the diagonal.

Since experiments on optimally-doped electron superconductors do find large
AFM correlation lengths \cite{Greven03} as well as $\omega =0$
single-particle excitations \cite{Armitage01} near the diagonal, the
pseudogap mechanism in this case should be the weak-coupling one $(
U\lesssim 6t)$~\cite{Kyung02, Markiewicz}. This value of $U$ is
smaller than, but not too different from, that necessary for a sizeable Mott gap
at half-filling. This may be understood as follows. The contribution to the
value of $U$ that comes from simple Thomas-Fermi screening scales like
$(\partial\mu/\partial n)$. Fig.~\ref{FIG:mu_vs_n} clearly
shows that this quantity, beginning at $U>4t$, is smaller for electron-doped than
for hole-doped systems, demonstrating the internal consistency of a picture
where the value of $U$ decreases as one goes from the hole- to the
electron-doped systems. Ref.~\cite{Kyung02} presents additional arguments
for a smaller $U$.

To summarize, we illustrated two ways in which a Fermi liquid can be
destroyed by a pseudogap and found that a unified picture of $A(\mathbf{k}%
,\omega )$ in the cuprates emerges from the $t-t^{\prime }-t^{\prime \prime
}-U$ model if we allow $U$ to decrease as the concentration of electrons
increases.


\begin{acknowledgments}
We are indebted to B. Kyung, V. Hankevych, K. Le Hur and J. Hopkinson for
useful discussions. The present work was supported by the Natural Sciences
and Engineering Research Council (NSERC) of Canada, the Fonds qu\'eb\'ecois
de la recherche sur la nature et les technologies (FQRNT), the Canadian
Foundation for Innovation, and the Tier I Canada Research Chair Program
(A.-M.S.T.).
\end{acknowledgments}



\end{document}